%% file: main.tex
\pdfoutput=1
\documentclass[a4paper,11pt]{article}
\usepackage{jcappub}
\hypersetup{hypertexnames=false}
\usepackage{booktabs}
\usepackage{float}  
\graphicspath{{pic/}}

\title{High-energy neutrino signatures of embedded GRB jets in AGN disks: a dynamic jet-propagation framework}

\author[a,b]{Wei-Cheng Long}
\author[a,b,1]{and Yun-Wei Yu,\note{Corresponding author.}}

\affiliation[a]{Institute of Astrophysics, Central China Normal University, Wuhan 430079, China}
\affiliation[b]{Laboratory for Compact Object Astrophysics and Astronomical Technology, Central China Normal University, Wuhan 430079, China}

\emailAdd{wclong@mails.ccnu.edu.cn}
\emailAdd{yuyw@ccnu.edu.cn}

\abstract{
Relativistic jets embedded in active galactic nucleus (AGN) accretion disks are promising high-energy neutrino sources, but their emission is often estimated from a single representative jet state. We develop a time-dependent framework that follows jet-head propagation, evolving reverse-shock conditions, and particle cooling until the jet chokes or breaks out, and apply  it to the SG and TQM disk models. For the representative choked cases, neutrino emission is dominated by the high-dissipation phase near jet stalling, allowing a stalling-state approximation to reproduce the trajectory-integrated, detector-weighted event yield within approximately $14\%$. In breakout cases, however, rapid jet-head acceleration across steep disk-density gradients suppresses reverse-shock dissipation and can cause single-state estimates to overpredict the fluence, even after accounting for the available energy budget. Full trajectory integration also reshapes the high-energy spectral tail and produces distinct detectability patterns across SMBH mass and disk radius for the two disk models. Some lower-density outer-disk cases develop harder tails extending into the 10--100 PeV range, motivating future ultra-high-energy neutrino searches. Resolving jet propagation dynamics is therefore indispensable for evaluating embedded transients across AGN disk environments and avoiding systematic biases in multi-messenger modeling.
}

\keywords{High-energy neutrinos, active galactic nuclei, relativistic jets, gamma-ray bursts, accretion disks}

\begin{document}
\maketitle
\flushbottom

\section{Introduction}%

Accretion disks around supermassive black holes (SMBHs) in active galactic nuclei (AGNs) may host substantial populations of stars and compact objects, which can be built up through in situ star formation and the capture of objects from surrounding nuclear star clusters, while gas-driven migration can redistribute embedded objects within the disk~\cite{artymowicz1993, goodman2003, mckernan2012, mckernan2020, tagawa2020}.  Interaction with the dense disk gas can alter the mass growth and orbital evolution of embedded stars and compact objects relative to their counterparts in lower-density environments~\cite{cantiello2021, wang2021, tagawa2020}, potentially enhancing the rates of massive-star collapse~\cite{cantiello2021, wang2021} and compact-object mergers and collisions~\cite{cheng1999a, bartos2017, mckernan2020, tagawa2020, zhu2024a}.  As a result, AGN disks are promising sites for frequent launching of relativistic jets reminiscent of those accompanying gamma-ray bursts~(GRBs)~\cite{woosley2006, berger2014}.  Relativistic jets may also be launched by isolated accreting black holes~\cite{tagawa2023a} and by merger-remnant black holes embedded in the AGN disks~\cite{wang2021a, tagawa2023b}.

To produce directly observable prompt emission, a GRB jet must first penetrate the surrounding progenitor material~\cite{bromberg2011, gottlieb2022b}. A GRB jet launched within an AGN disk must additionally traverse the dense disk gas and break out of the disk. Studies of jet--disk interactions by Zhu et al.~\cite{zhu2021a, zhu2021b} and Zhang et al.~\cite{zhang2024} have shown that such jets are often choked within the dense AGN disks and can successfully break out of the disk only if the central engine remains active for a sufficiently long time. Therefore, compared with ordinary GRBs, the direct jet emission such as GRB-like prompt emission may be strongly suppressed, while the observable electromagnetic counterpart may instead be dominated by a relatively faint transient powered by the jet-driven cocoon~\cite{zhu2021a}. Nevertheless, relativistic jets have long been considered promising sources of high-energy neutrinos~\cite{waxman1997, murase2006a, yu2008, hummer2012, baerwald2015, biehl2018a}, although no conclusive association with the prompt emission of ordinary GRBs has yet been established~\cite{aartsen2017, abbasi2024a}. Hidden GRB jets in AGN disks therefore provide a possible channel for producing high-energy neutrinos with potentially detectable fluences in the TeV--PeV range~\cite{zhu2021b}, since neutrino production may remain efficient despite the strong suppression of the accompanying electromagnetic emission.%

However, existing estimates of high-energy neutrino emission from choked jets in AGN disks do not account for the full jet propagation history. The pioneering calculation by Zhu et al.~\cite{zhu2021b} adopted a single-state approximation tailored for choked jets, in which the disk density and reverse-shock microphysics were evaluated at a characteristic jet-stalling radius. By construction, this stalling-state framework is designed for embedded choked jets and is not intended to describe jets that successfully break out of the disk. The neutrino fluence instead depends on the evolving dissipation power, photomeson efficiency, maximum proton energy, and cooling of secondary mesons and muons. Different parts of the time-integrated spectrum can consequently be dominated by different stages of the propagation. It is therefore essential to systematically evaluate the validity, precision, and applicable physical regimes of such single-state approximations relative to full time-integrated propagation. More recently, Zhang et al.~\cite{zhang2025b} showed, in the context of ordinary relativistic jets, that integrating neutrino production over the jet expansion history can significantly modify the resulting neutrino fluence relative to the commonly adopted instantaneous treatment. This work presents a time-dependent calculation that follows the full propagation histories of hidden jets in AGN disks and evaluates the precision and scope of applicability of single-state approximations across both choked and breakout regimes.

This paper is organized as follows. Section~\ref{sec:Models} describes the jet propagation dynamics in AGN disks and the calculation of time-dependent high-energy neutrino production. Section~\ref{sec:single} presents the neutrino emission from individual relativistic jets in AGN disks, as well as the parameter dependences and detectability. The main findings and their implications for future observations are presented in section~\ref{sec:discussion}. 

\section{Models}%
\label{sec:Models}
\subsection{Jet head dynamics and the neutrino production environment}%
The head region of a propagating jet is confined by a forward and reverse shock, which propagate into the ambient disk material and the unshocked relativistic jet, respectively. We consider a relativistic jet powered by a central engine, with an isotropic-equivalent kinetic energy $E_{\mathrm{iso}}$ and an engine duration $t_{\mathrm{dur}}$ after the jet has emerged from the stellar envelope. For one side of an axisymmetric top-hat jet with injection half-angle $\theta_0$, the constant physical luminosity is $L_{\mathrm j}=(E_{\mathrm{iso}}/t_{\mathrm{dur}})(1-\cos\theta_0)/2$. The jet has an initial Lorentz factor of $\Gamma_{\mathrm{j}}$, and its propagation through the AGN disk is initialized at the surface of a progenitor star located in the disk midplane. The stellar radius is set to $R_{\star} = 10^{10.5}\,\mathrm{cm}$. Following refs.~\cite{matzner2003,bromberg2011,zhang2024}, the jet-head velocity is then determined as
\begin{equation}
   \beta_{\mathrm{h}}(t) \approx \frac{\beta_{\mathrm{j}}}{1+\tilde{L}(t)^{-1/2}}. \label{eq:betah}
\end{equation}
Here the dimensionless jet-head luminosity is defined as
\begin{equation}
    \tilde L(t) \equiv \frac{{L}_{{\rm{j}}}(t_{\mathrm{e}})}{{\rm\Sigma}_{{\rm{h}}}(t){\rho }_{\rm{d}}(z_{\mathrm{h}}){c}^{3}}, \quad 0 \leq t_{\mathrm{e}}\le t_{\mathrm{dur}}.\label{eq:tildeL}
\end{equation}
Taking $t=0$ when the jet head reaches the stellar surface $z=R_\star$, the central-engine time that has reached the head is $t_{\mathrm{e}}(t)=t-\frac{z_{\mathrm h}(t)-R_\star}{\beta_{\mathrm j}c}$. Thus $t$ is the source-frame propagation time of the jet head, whereas $t_{\mathrm{e}}$ is the retarded engine time reaching the head. All time axes in this work represent this source-frame propagation time. Cosmological redshift and time dilation are neglected for the fiducial $100\,\mathrm{Mpc}$ benchmark, consequently, no factor of $1+z$ is applied to the plotted times or energies. The head cross section is evaluated from the instantaneous cocoon pressure as $\Sigma_{\rm h}=\min\{\pi(\theta_0z_{\rm h})^2, L_{\rm j}\theta_0^2/(4P_{\rm c}c)\}$, with $P_{\mathrm{c}}$ the cocoon pressure~\cite{bromberg2011,zhang2024}.

Following ref.~\cite{zhang2024}, we adopt a Gaussian vertical gas-density profile for the AGN disks. The profile is determined by the adopted disk model as a function of the SMBH mass $M_{\rm SMBH}$ and radial distance $a$ from the SMBH. The latter is expressed in units $r_{\rm g}\equiv GM_{\rm SMBH}/c^2$. The SG~\cite{sirko2003} and the TQM~\cite{thompson2005} models are considered, and the effective disk surface is defined at $z=\hat z$, beyond which the gas density transitions to a constant ambient interstellar-medium density of $\rho_{\mathrm{ism}} = 10^{-25} \mathrm{g}\,\mathrm{cm}^{-3}$.  We assume an initially unperturbed disk environment and neglect low-density cavities that may be excavated around some embedded compact-object binaries by pre-merger accretion-driven outflows~\cite{kimura2021, chen2023}. In our fiducial setup, the jet has emerged from the progenitor envelope, with a bulk Lorentz factor of the unshocked jet $\Gamma_j = 300$, an engine duration $t_{\mathrm{dur}} = 10^{1.5}\, \mathrm{s}$, and an isotropic-equivalent kinetic energy $E_{\mathrm{iso}} = 10^{53}\, \mathrm{erg}$. We further adopt an initial half-opening angle of $\theta_0 = 5^\circ$. 

In the dense AGN disk, the forward shock is typically radiation mediated before breakout, so efficient particle acceleration at the forward shock is suppressed~\cite{levinson2008, levinson2020, murase2013}. For the reverse shock, the comoving proton number density of the unshocked jet and its relative Lorentz factor with respect to the head are $n'_j=L_{\mathrm{j}}/(\Sigma_h\Gamma_j^2\beta_jm_pc^3)$ and $\bar\Gamma_h=\Gamma_j\Gamma_h(1-\beta_j\beta_h)$, respectively, where the local head cross section $\Sigma_h$ is used, also after collimation. The reverse shock is initially a radiation-mediated shock (RMS) but can become collisionless if the Thomson optical depth of the unshocked jet satisfies
\begin{equation}\label{eq:tauT}
\tau_T=\frac{\sigma_Tn'_jz_h}{\Gamma_j}
\leq\min\!\left[1,\frac{0.1\bar\Gamma_h}{1+2\ln\bar\Gamma_h^2}\right].
\end{equation}
Here $\sigma_{\mathrm T}$ is the Thomson cross section and the logarithmic term accounts for pair loading. We evaluate this condition along the jet-head trajectory and include neutrino production only during stages in which the reverse shock is collisionless. The jet-head dynamics determines the reverse-shock conditions for proton injection, while the dense, optically thick disk maintains a thermalized radiation field until choking or breakout. Together, these ingredients define the time-dependent environment for neutrino production. The shock-breakout position $z_{\mathrm{bo}}$ is defined by the photon diffusion condition ahead of the forward shock, $\tau(z_{\mathrm{bo}})\simeq c/v(z_{\mathrm{bo}})$, and the corresponding shock breakout time is $t_{\mathrm{bo}}$. Because the luminosity supplied to the jet head is determined by the engine time reaching the head, we define the minimum source-engine duration required for breakout along a continuously powered trajectory as
\begin{equation}\label{eq:tcr}
t_{\rm req}=\int_{0}^{t_{\mathrm{bo}}}\left(1-\frac{\beta_h}{\beta_j}\right)dt.
\end{equation}
Since the jet is ultra-relativistic ($\Gamma_{\rm j} = 300$, so $\beta_{\rm j} \approx 1$), the relative velocity factor $1-\beta_{\rm h}/\beta_{\rm j}$ simplifies to $1-\beta_{\rm h}$, which serves as the dynamical factor governing the rate at which jet power is processed by the reverse shock. This is the finite-$\beta_{\rm j}$ engine-clock definition used for the breakout/choking classification. We define the dimensionless engine-clock ratio $\mathcal{R}_{\rm dur} \equiv t_{\rm req}/t_{\rm dur}$: shock breakout occurs if $\mathcal{R}_{\rm dur} \le 1$, and the jet is choked within the disk if $\mathcal{R}_{\rm dur} > 1$. We denote the end of the embedded propagation by $t_{\rm end}$: it is the optical-depth breakout time for a successful jet and the finite-engine choking time otherwise.

\begin{figure}[H]
    \centering
    \includegraphics[width=0.6\textwidth]{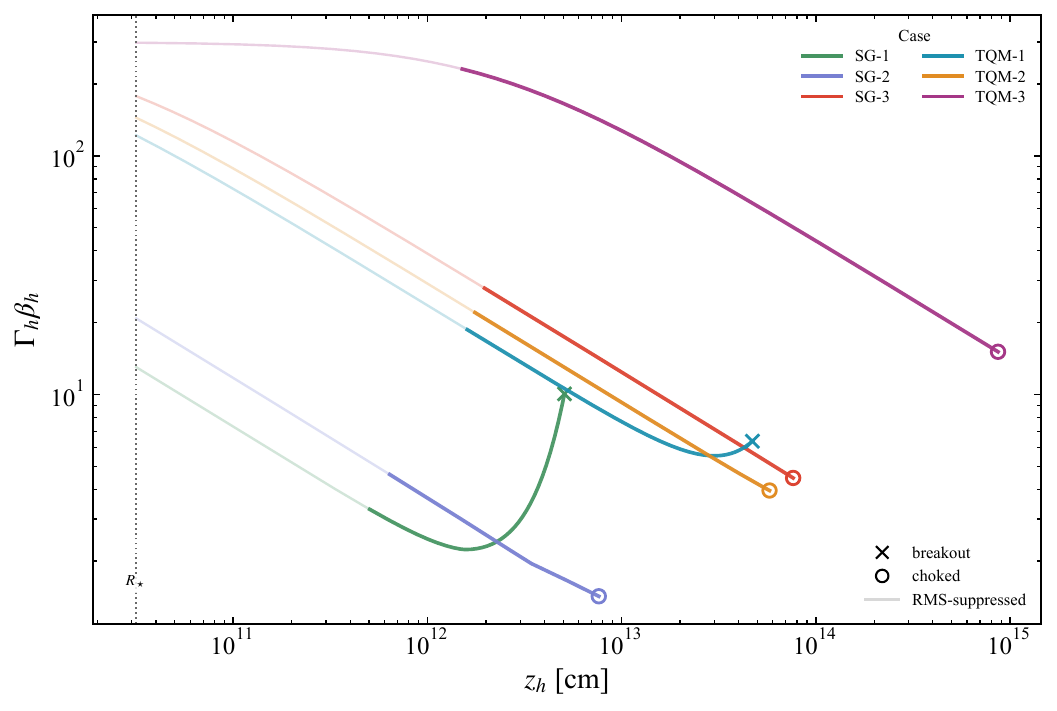}
    \caption{Dynamical evolution of the six representative jet heads as a function of height $z_{\mathrm h}$. Colors encode the six representative cases ($\mathrm{SG}\text{--}1\dots 3$ and $\mathrm{TQM}\text{--}1\dots 3$). Open circles mark choking, whereas crosses mark successful breakout. Faint curve segments fail the state-dependent RMS acceleration condition; fully colored segments enter the observable neutrino fluence. The corresponding disk parameters are listed in table~\ref{tab:disk_parameters_jetprop}.}
    \label{fig:zhvsGmbt}
\end{figure}

Figure~\ref{fig:zhvsGmbt} shows the jet-head proper velocity $\beta_{\mathrm{h}}\Gamma_{\mathrm{h}}$ as a function of the jet head position $z_{\mathrm{h}}$ for different AGN disk models, with open circles marking choking positions and crosses marking successful breakout. Owing to the stratified disk density, the jet head initially decelerates in the dense disk interior and can subsequently accelerate as it approaches the disk surface, see~\cite{zhang2024} for more details. These trajectories provide the time-dependent conditions for the neutrino production calculation. Unlike treatments that approximate the evolution of an embedded or choked jet using a single characteristic stage, our model follows the evolving shock conditions throughout the full propagation trajectory. To cover the diverse jet propagation outcomes across the parameter space, we select three representative parameter combinations of SMBH mass and orbital radius for each disk model based on their distinct engine-clock ratios $\mathcal{R}_{\rm dur}$ (table~\ref{tab:disk_parameters_jetprop}), spanning both the breakout ($\mathcal{R}_{\rm dur} \le 1$) and choked ($\mathcal{R}_{\rm dur} > 1$) regimes. The outcome is primarily controlled by this engine-clock ratio rather than by the laboratory propagation time alone.

\begin{table}[tbp]
    \centering
    \input{pic/disk_parameters_sixproto.tex}
    \caption{AGN-disk parameters for the six representative jet-propagation trajectories shown in figure~\ref{fig:zhvsGmbt}. The case label encodes the underlying disk prescription (SG or TQM). The tabulated coordinates are $\log_{10}(M_{\rm SMBH}/M_\odot)$ and $\log_{10}(a/r_{\rm g})$, where $a$ is the orbital radius and $r_{\rm g}=GM_{\rm SMBH}/c^2$ is the gravitational radius of the central SMBH. The quoted midplane gas density $\rho_0$, opacity $\kappa$, and scale height $H$ are evaluated from the corresponding disk model. The last column gives the engine-clock ratio $\mathcal{R}_{\rm dur} \equiv t_{\rm req}/t_{\rm dur}$ relative to the adopted duration $t_{\rm dur}=10^{1.5}\,\mathrm{s}$; values $\mathcal{R}_{\rm dur} \le 1$ denote successful shock breakout.}
    \label{tab:disk_parameters_jetprop}
\end{table}

\subsection{Time-dependent neutrino production}%
\label{sec:pronu}
We calculate time-dependent neutrino emission from non-thermal protons accelerated at the reverse shock whenever eq.~\eqref{eq:tauT} is satisfied. All local shock parameters and target radiation fields are evaluated dynamically along the jet-head trajectory. 

The comoving internal energy densities behind the forward and reverse shocks are given by relativistic shock jump conditions~\cite{blandford1976, sari1995},
\begin{equation} \label{eq:shock jump}
\begin{array}{rcl}
    e_{\rm f}^{\prime} & = & 4\Gamma_{\rm h}(\Gamma_{\rm h}-1)n_{\rm a} m_p c^{2}, \\
    e_{\rm r}^{\prime} & = & 4\bar{\Gamma}_{\rm h}(\bar{\Gamma}_{\rm h}-1)n^{\prime}_{\rm j} m_p c^{2},
\end{array}
\end{equation}
where $n_a=\rho_d/m_p$ is the ambient disk proton number density. We parameterize the thermal radiation and magnetic energy densities in the reverse-shock region using microphysical energy-partition fractions, $u'_\gamma = \varepsilon_e e'_r$ and $u'_B = B'^2/(8\pi) = \varepsilon_B e'_r$, with fiducial values $\varepsilon_e=0.1$ and $\varepsilon_B=0.1$. Prior to shock breakout, the reverse-shock photon field is assumed to be thermalized at a comoving temperature $k_{\rm B}T'_r=(15\hbar^3c^3\varepsilon_e e'_r/\pi^2)^{1/4}$, yielding a characteristic photon energy $\epsilon'_\gamma = 2.7 k_{\rm B}T'_r$ and photon number density $n'_\gamma = \varepsilon_e e'_r / \epsilon'_\gamma$.

Protons are accelerated at the reverse shock via first-order Fermi acceleration with an injected power-law spectrum $dn_{p}^\prime / d\epsilon_p^\prime \propto {\epsilon_p^\prime}^{-s}$ ($s=2$). The minimum proton energy is $\epsilon^\prime_{p,\min} = \bar{\Gamma}_{\mathrm{h}} m_{p}c^2$, here $\bar\Gamma_{\rm h} \approx \Gamma_{\rm j}\tilde{L}^{-1/4}/\sqrt{2}$. The maximum proton energy $\epsilon'_{p,\max}$ is determined by balancing the acceleration timescale $t'{}_{p,\mathrm{acc}} = \eta_{\rm acc}\epsilon'_{p} / (eB'c)$ against the total cooling timescale, where $B'=\sqrt{8\pi\varepsilon_B e'_r}$ and $\eta_{\rm acc}=1$ denotes the Bohm-limit benchmark. Accelerated protons cool through $pp$ and $p\gamma$ hadronic interactions, Bethe--Heitler (BH) pair production, synchrotron radiation, inverse Compton (IC) scattering, and adiabatic expansion~\cite{zhu2021b}.

Neutrinos are produced through the decay of secondary pions, kaons, and daughter muons. The channel-resolved neutrino production efficiency is quantified by energy-dependent suppression factors $\zeta_{p,\sup}$ and $\zeta_{i,\sup}$ ($i = \pi,\;\mathrm{K},\;\mu_\pi,\;\mu_{\rm K}$), following the energy-partition prescription of refs.~\cite{denton2018, zhu2021b}. The fraction of proton energy losses converted into $pp$ and $p\gamma$ interactions is
\begin{equation}
{\zeta }_{p,\;\mathrm{sup} }({\epsilon }_{{\nu }_{i}})=\frac{t_{p,\;{pp}}^{\prime-1}+t_{p,\;p\gamma }^{\prime-1}}{t_{p,\; \mathrm{cool}}^{\prime -1}},
\end{equation}
where $t_{p,\;\mathrm{cool}}^{\prime-1}$ accounts for all hadronic, radiative, and expansion cooling losses. For secondary mesons and muons, cooling prior to decay is accounted for by
\begin{equation}
{\zeta }_{i,\;\mathrm{sup} }({\epsilon }_{{\nu }_{i}})=\frac{t _{i,\;\mathrm{dec}}^{\prime -1}}{{t}_{i,\;\mathrm{dec}}^{\prime -1}+t_{i,\;\mathrm{had}}^{\prime -1} + t_{i,\;\mathrm{syn}}^{\prime-1}+t_{i,\;\mathrm{IC}}^{\prime -1}+t_{i,\;\mathrm{adi}}^{\prime -1}}.
\end{equation}

Crucially, in our framework all local shock conditions, target fields, and particle cooling rates evolve along the jet-head trajectory. Combining reverse-shock dissipation power, proton interaction efficiency, and secondary decay probability, the unoscillated channel-resolved neutrino energy flux at the observer is
\begin{equation}\label{eq:flux-time}
    {\epsilon }_{\nu_i}^{2}\phi_{{\nu}_{i}}^{\mathrm{0}}(\epsilon_{\nu_i},\,t)=\frac{{N}_{i} L_{\mathrm{j}}(t_{\mathrm{e}})(1- \beta_{\mathrm{h}}(t))\zeta_{p,\sup }({\epsilon }_{{\nu }_{i}}, t){\zeta }_{i,\sup }({\epsilon }_{{\nu }_{i}}, t)}{\Omega_{0} {D}_{L}^{2}\ln({\epsilon }_{p,\max }^{{\prime} }(t)/{\epsilon }_{p,\min }^{{\prime} }(t))},
\end{equation}
where superscript 0 denotes unoscillated quantities, $\Omega_0=2\pi(1-\cos\theta_0)$ is the injection solid angle, and $N_i$ is the channel normalization~\cite{denton2018}. The factor $L_{\mathrm{j}}(1-\beta_{\mathrm{h}})$ quantifies the instantaneous rate of jet energy processed by the reverse shock. The total unoscillated time-integrated fluence is obtained by integrating over the active propagation phase, which gives
\begin{equation}\label{eq:fluence-time}
    {\epsilon }_{{\nu }_{i}}^{2}{F}_{{\nu}_{i}}^{\mathrm{0}}(\epsilon_{\nu_i})=\int_{0}^{t_{\mathrm{end}}}{\epsilon }_{{\nu }_{i}}^{2}{\phi}_{{\nu }_{i}}^{\mathrm{0}}(t)dt.
\end{equation}

For comparison, we also evaluate fluences using standard single-state baselines. For choked jets, we evaluate the stalling-state approximation of ref.~\cite{zhu2021b}, where shock conditions and cooling efficiencies are fixed at the characteristic stalling radius $r_{\rm stall}$:
\begin{equation}\label{eq:fluence-instant}
    {\epsilon }_{{\nu }_{i}}^{2}{F}_{{\nu}_{i},\,\mathrm{stat}}^{\mathrm{0}}=\frac{{N}_{i}{E}_{\mathrm{iso}}{\zeta }_{p,\sup }({\epsilon }_{{\nu }_{i}}){\zeta }_{i,\sup }({\epsilon }_{{\nu }_{i}})}{4\pi {D}_{L}^{2}\ln({\epsilon }_{p,\max }^{{\prime} }/ {\epsilon }_{p,\min }^{{\prime }})}.
\end{equation}
For the breakout trajectories, we additionally apply the same uniform-density stalling-state prescription as a controlled formal reference, without imposing the disk-height restriction $r_{\rm stall}<H$ assumed in the ref.~\cite{zhu2021b}. We then evaluate a separate pre-breakout approximation at the actual state immediately before breakout, and replacing $E_{\rm iso}$ with $E_{\rm iso}\mathcal{R}_{\rm dur}$ to account for the engine energy that reaches the head before breakout.

Combining electron- and muon-decay channels ($F_{\nu_e+\bar\nu_e}^{\rm 0}=F_{\mu\pi}+F_{\mu_K}$, $F_{\nu_\mu+\bar\nu_\mu}^{\rm 0}=F_{\pi}+F_K+F_{\mu\pi}+F_{\mu_K}$, and $F_{\nu_\tau+\bar\nu_\tau}^{\rm 0}=0$), the observable flavor fluences at Earth are obtained via standard oscillation-averaged PMNS matrix mixing~\cite{bustamante2019c,esteban2024}:
\begin{equation}
    \label{eq:oscillation}
F_{\nu_\beta+\bar\nu_\beta}^{\oplus} = \sum_\alpha P_{\alpha\beta} F_{\nu_\alpha+\bar\nu_\alpha}^{\mathrm 0},
\end{equation} 
where $P_{\alpha\beta}=\sum_i |U_{\alpha i}|^2|U_{\beta i}|^2$ is the transition probability. We construct $U$ using the NuFIT~6.1\footnote{NuFIT:~\url{http://www.nu-fit.org/}} normal-ordering global fit parameters ($\sin^2\theta_{12}=0.3088$, $\sin^2\theta_{23}=0.470$, $\sin^2\theta_{13}=0.02248$, $\delta_{\rm CP}=212^\circ$)~\cite{esteban2024}.

\section{Results} \label{sec:single}
\subsection{Jet dynamics and time-resolved emission}%
All six representative trajectories undergo one transition from an RMS to a collisionless reverse shock, with no subsequent re-entry into the RMS regime. The reverse shock transitions to a collisionless state shortly after emerging from the stellar envelope. Because proton acceleration is suppressed during the RMS phase, these initial intervals are excluded from the integration, and only the collisionless portions contribute to the observable neutrino signal.

As listed in table~\ref{tab:disk_parameters_jetprop}, the six representative cases sample three distinct dynamical regimes of jet propagation:
(i) \emph{Breakout trajectories} (SG-1 and TQM-1, with $\mathcal{R}_{\rm dur} \le 1$), where the jet head penetrates the disk surface before engine exhaustion;
(ii) \emph{Moderately choked trajectories} (SG-2 and TQM-2, with $\mathcal{R}_{\rm dur} \sim 10\text{--}50$), where the jet head chokes deep inside dense interior gas; and
(iii) \emph{Deeply choked trajectories} (SG-3 and TQM-3, with $\mathcal{R}_{\rm dur} \gg 10^3$), where the jet head chokes after prolonged embedded propagation through extended gas.

\begin{figure}[tpb]
    \centering
    \includegraphics[width=0.8\textwidth]{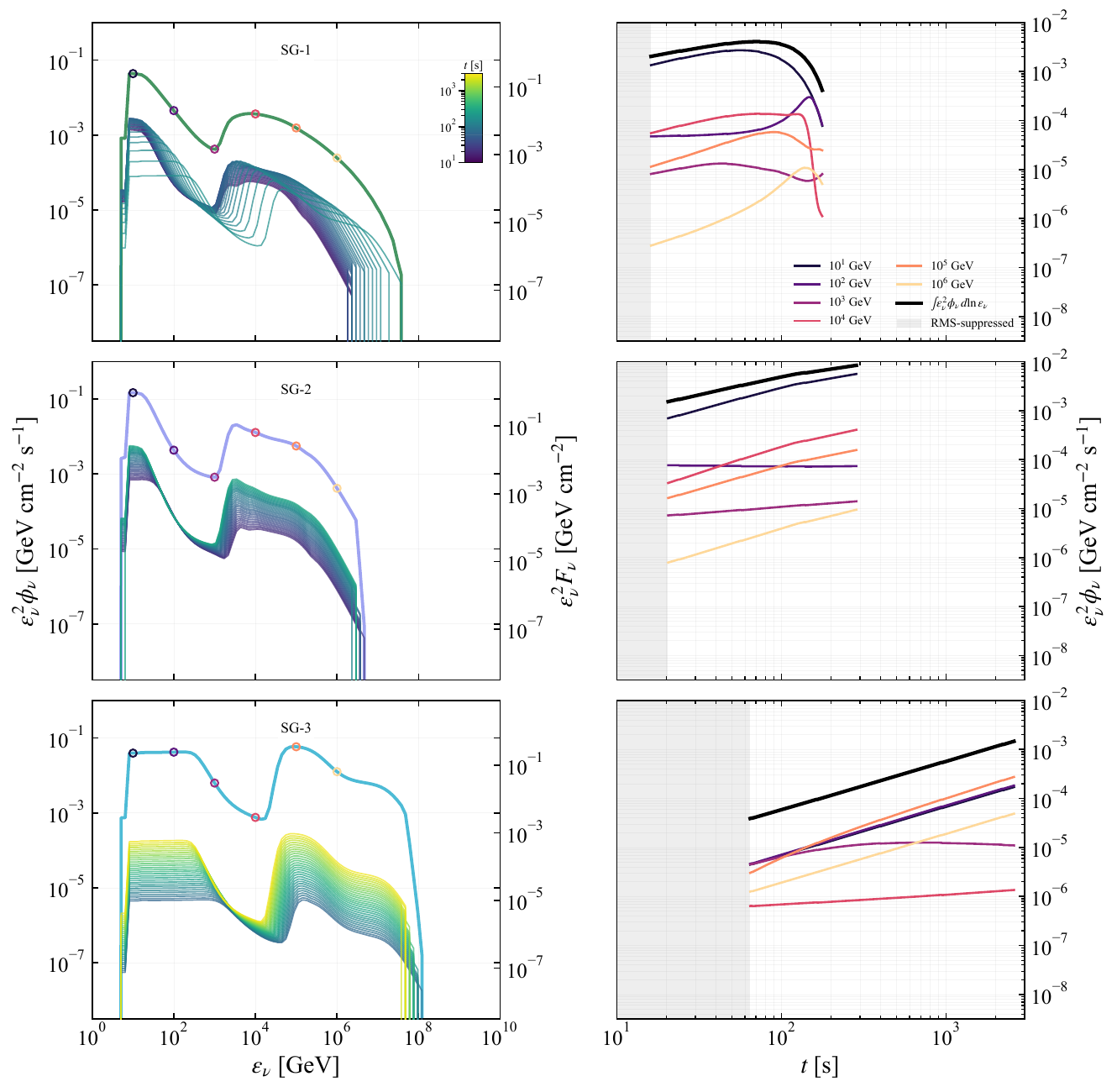}
    \caption{Time evolution of the unoscillated, all-flavor neutrino and antineutrino emission in the SG disk model; panel titles give the case name. The top row shows a successful breakout, whereas the middle and bottom rows show finite-engine choked jets. Left panels show instantaneous \(E_\nu^2\phi_\nu\) spectra, colored by time, together with the time-integrated spectrum. Right panels show light curves at fixed neutrino energies. Radiation-mediated portions of each trajectory are excluded from the observable fluence.}
    \label{fig:nuslsg}
\end{figure}

\begin{figure}[htpb]
    \centering
    \includegraphics[width=0.8\textwidth]{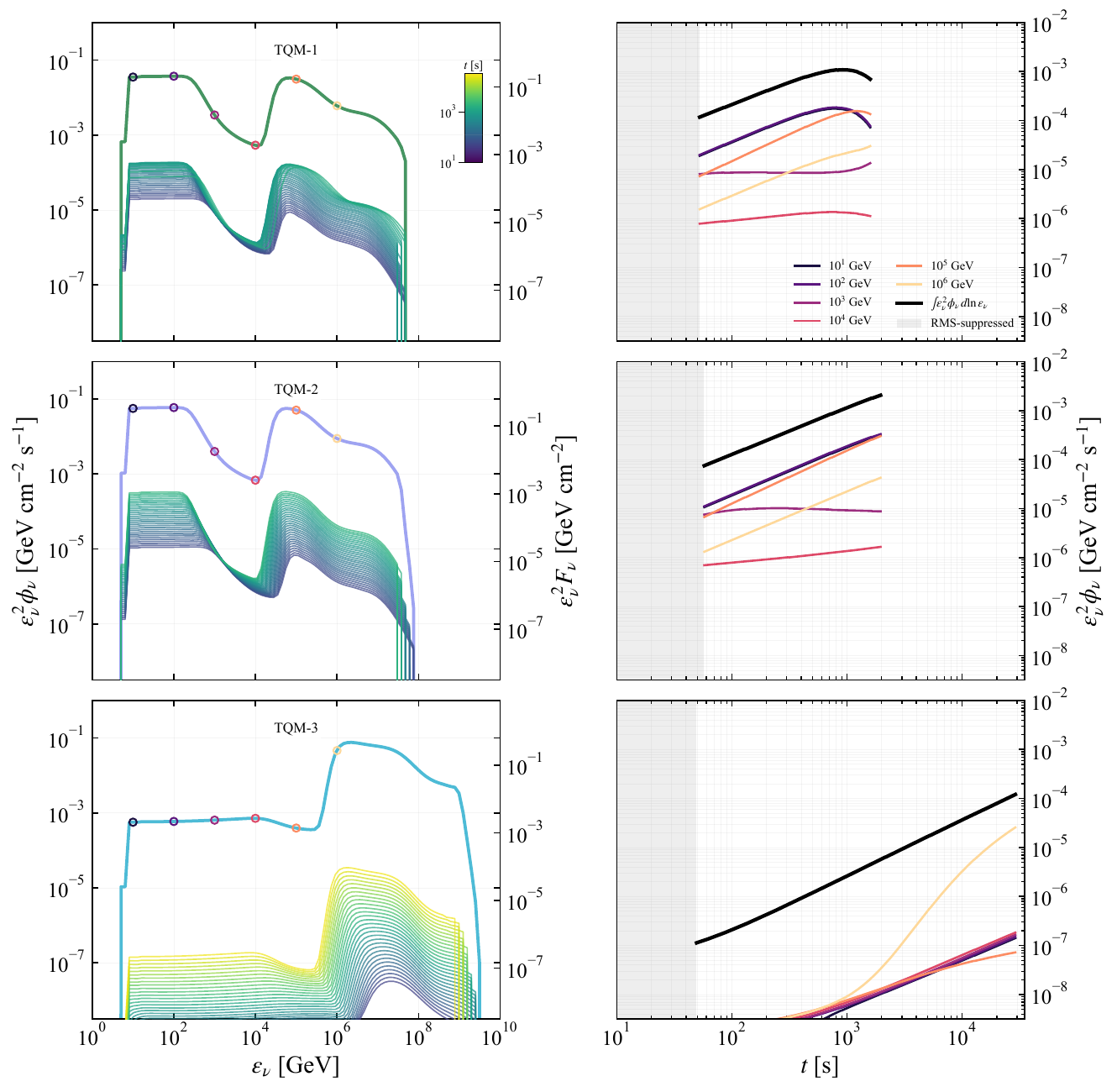}
    \caption{Time evolution of the unoscillated, all-flavor neutrino and antineutrino emission in the TQM disk model. The top row shows a successful breakout (TQM-1), whereas the middle and bottom rows show finite-engine choked jets (TQM-2 and TQM-3); the remaining conventions are the same as in figure~\ref{fig:nuslsg}.}
    \label{fig:nusltqm}
\end{figure}

The time-resolved instantaneous spectra and fixed-energy light curves for the SG and TQM models are shown in figures~\ref{fig:nuslsg} and \ref{fig:nusltqm}, respectively. The breakout trajectories (SG-1 and TQM-1) display a distinct time-domain signature: the instantaneous flux increases during head deceleration and turns over as the head accelerates toward the surface. For choked trajectories, the light curves rise continuously or flatten toward the end of the embedded phase, terminating abruptly upon engine exhaustion without a breakout decline. In addition, the instantaneous spectra evolve in both overall normalization and high-energy cutoff, reflecting the non-stationary shock conditions, for TQM-3, the ultra-dilute environment minimizes secondary meson and muon cooling, allowing the spectrum to extend to $\sim10^2\,\mathrm{PeV}$.

\begin{figure}[tbp]
  \centering
  \includegraphics[width=0.9\textwidth]{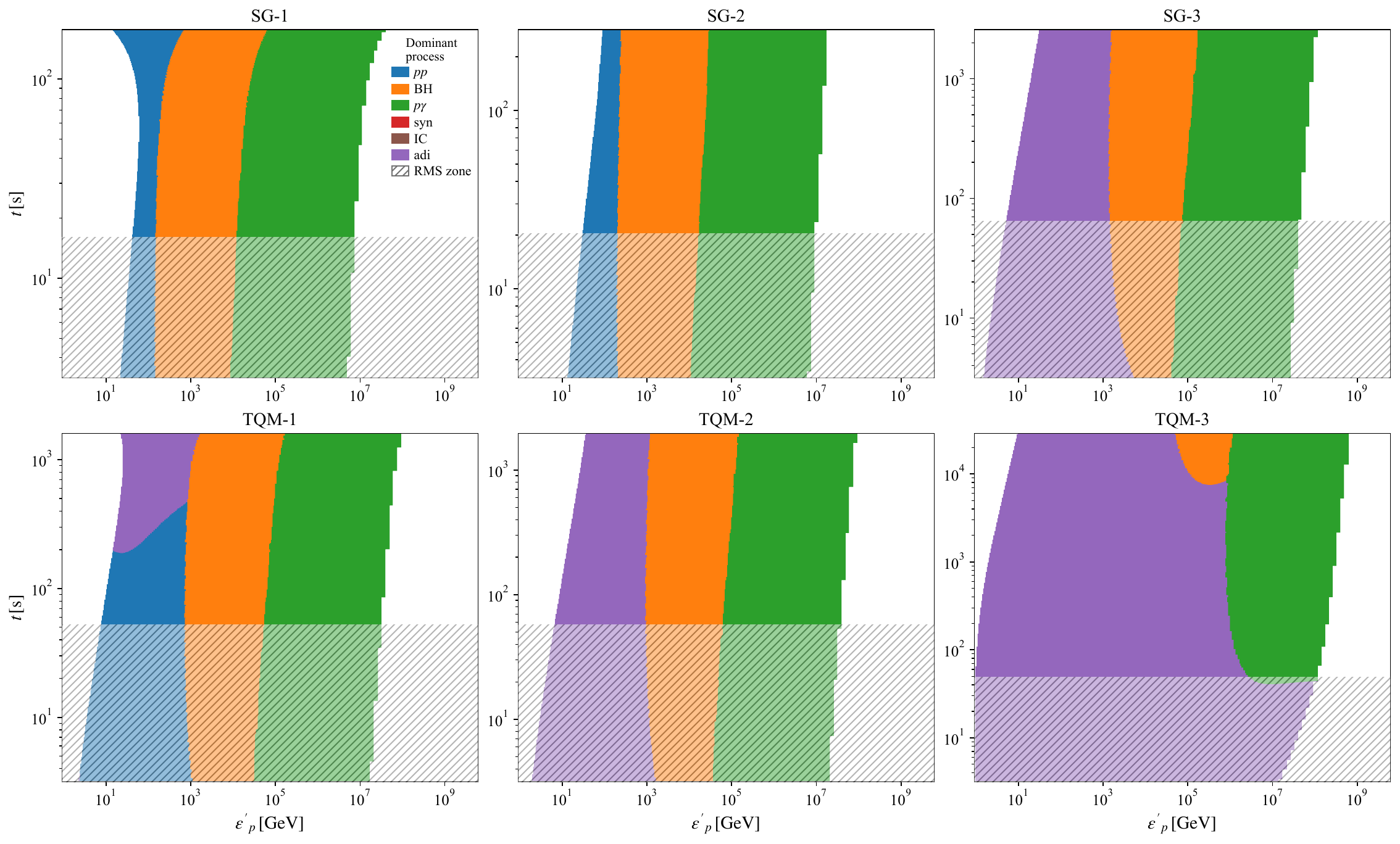}
  \caption{Dominant comoving proton cooling and interaction processes as functions of proton energy and propagation time across all six representative cases; panel titles indicate case names. Upper and lower rows correspond to the SG and TQM disk models, respectively. The color maps compare regions dominated by $pp$, $p\gamma$, BH pair production, synchrotron, IC, and adiabatic losses. The hatched regions labeled 'RMS zone' denote early radiation-mediated shock stages where collisionless particle acceleration is suppressed; these unaccelerated stages are excluded from neutrino production.}
  \label{fig:dominant_process_maps}
\end{figure}

Figure~\ref{fig:dominant_process_maps} maps the dominant comoving proton energy-loss processes as functions of proton energy and propagation time across all six trajectories. 
Across all cases, a common process hierarchy is established: $p\gamma$ photomeson interactions dominate at the highest proton energies, BH pair production occupies an intermediate energy band, and either $pp$ interactions or adiabatic expansion dominates at lower energies. Because the comoving photon density scales as $n'_\gamma \propto {e'_r}^{3/4}$, the ratio of photon-to-proton target densities scales as $n'_\gamma/n'_p \propto \bar\Gamma_h^{1/2}z_{\rm h}^{1/2}$ in the relativistic reverse-shock regime ($\bar\Gamma_{\rm h} \gg 1$), continuously boosting $p\gamma$ and BH pair production over $pp$ collisions as the jet head advances during deceleration. Meanwhile, the proton acceleration rate ($t'{}_{p,\mathrm{acc}}^{-1} \propto {e'_r}^{1/2}$) decays more slowly than photon density during head deceleration, pushing the maximum proton energy $\epsilon_{p,\max}'$ upward until late-stage adiabatic cooling takes over. 

\subsection{Time-integrated fluence and detector yield}%
\begin{figure}[tbp]
    \centering
    \includegraphics[width=0.48\textwidth]{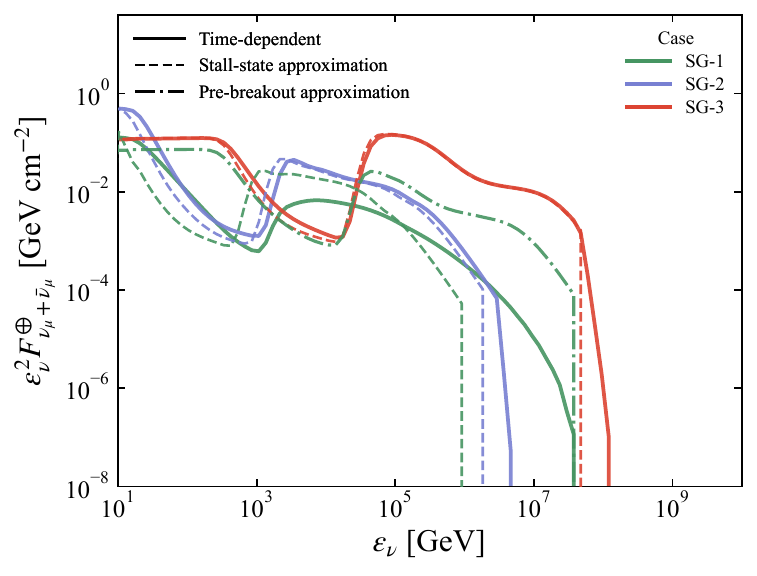}
    \includegraphics[width=0.48\textwidth]{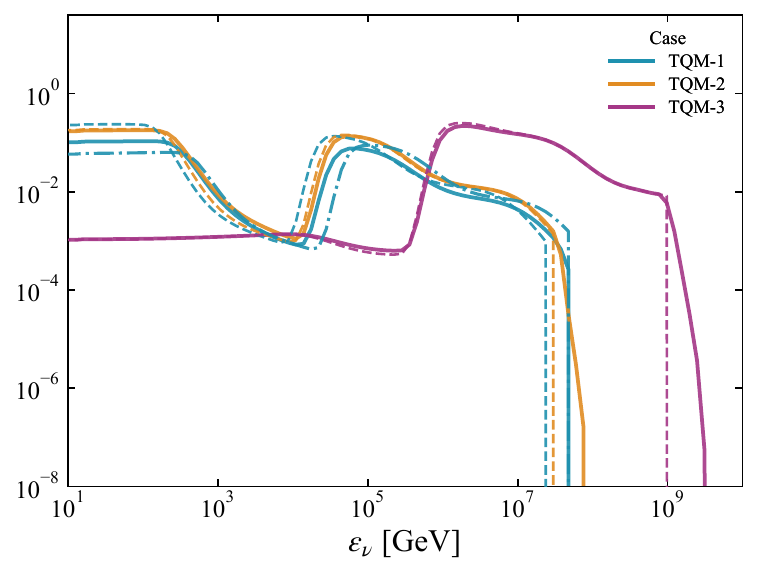}
    \caption{Earth-side $\nu_\mu+\bar{\nu}_\mu$ fluence from a single GRB jet embedded in an AGN disk. The left and right panels correspond to the SG and TQM disk models, respectively. Solid curves show the time-dependent calculation, and dashed curves show the stalling-state approximation. For the breakout cases SG-1 and TQM-1 only, dash-dotted curves show the pre-breakout approximation evaluated just before breakout and normalized by $E_{\rm iso}\mathcal{R}_{\rm dur}$. Colors identify the six representative cases.}
    \label{fig:esrFall}
\end{figure}

\label{sub:fluence_yields}
Figure~\ref{fig:esrFall} compares the time-dependent $\nu_\mu+\bar\nu_\mu$ fluence (solid lines) with the stalling-state approximation of ref.~\cite{zhu2021b} (dashed lines). For SG-1 and TQM-1, it also shows the pre-breakout approximation (dash-dotted lines). The fluences exhibit a low-energy component from $pp$ interactions and a high-energy component from $p\gamma$ interactions, separated by a depression associated with BH pair production. The time-dependent spectra generally extend to higher energies than the single-state curves, with the largest difference appearing for SG-1.

To translate the muon-flavor neutrino fluence at Earth into an expected detector event count,  we integrate 
\begin{equation}\label{eq:ndet}
N_{\rm det} = \int dE_\nu \, F_{\nu_\mu+\bar\nu_\mu}^{\oplus}(E_\nu)\,
A_{\rm eff}^{\nu_\mu+\bar\nu_\mu}(E_\nu),
\end{equation}
where $A_{\rm eff}^{\nu_\mu+\bar\nu_\mu}$ is the declination-averaged IC86 effective area for muon tracks~\cite{icecubecollaboration2021,collaboration2024}. At the benchmark distance of $D_L=100\,\mathrm{Mpc}$, the representative SG and TQM cases yield expected IC86 muon-track counts ranging from order $0.1$ to $1.4$, scaling as $N_{\rm det}\propto D_L^{-2}$ for otherwise fixed source parameters.

Table~\ref{tab:six_case_results} summarizes the detector event counts across the three physical model formulations. For all four choked trajectories, stalling-state approximations reproduce the full time-integrated event counts reliably (within $\approx 14\%$). This remarkable agreement arises because the reverse-shock dissipation power scales as $1-\beta_{\rm h}$, which reaches its peak when the jet head decelerates to its minimum velocity near the stalling radius deep inside the disk. Because emission from this peak shock dissipation phase dominates the integrated yield, evaluating a stalling-state approximation captures the bulk of the neutrino flux accurately.

For breakout trajectories, however, jet propagation extends into the outer disk layers where the jet head accelerates toward breakout. During this outward phase, the proton maximum energy $\epsilon_{p,\max}'$ continuously increases while the shock-processed power factor $1-\beta_{\rm h}$ drops. In TQM disks with gentle density gradients, evaluating the pre-breakout approximations with energy normalization $E_{\rm iso}\mathcal{R}_{\rm dur}$ brings the yield into close agreement with the time-dependent result (within $\approx 9\%$). In SG disks with steep density gradients, late-stage acceleration drives rapidly suppresses the shock-processed power, $1-\beta_{\rm h}\to 0$, an evolution that cannot be captured by a single state, leaving even the pre-breakout approximation a factor of $2.6$ above the true time-integrated yield. While energy truncation adjusts the total energy budget, it cannot correct errors driven by such dynamic dissipation evolution. Full trajectory integration is therefore indispensable for resolving this interplay between peak-dissipation choking phases and breakout acceleration.

\input{pic/six_case_results.tex}

\subsection{Parameter space and detection}%
\begin{figure}[tbp]
    \centering
    \includegraphics[width=0.9\textwidth]{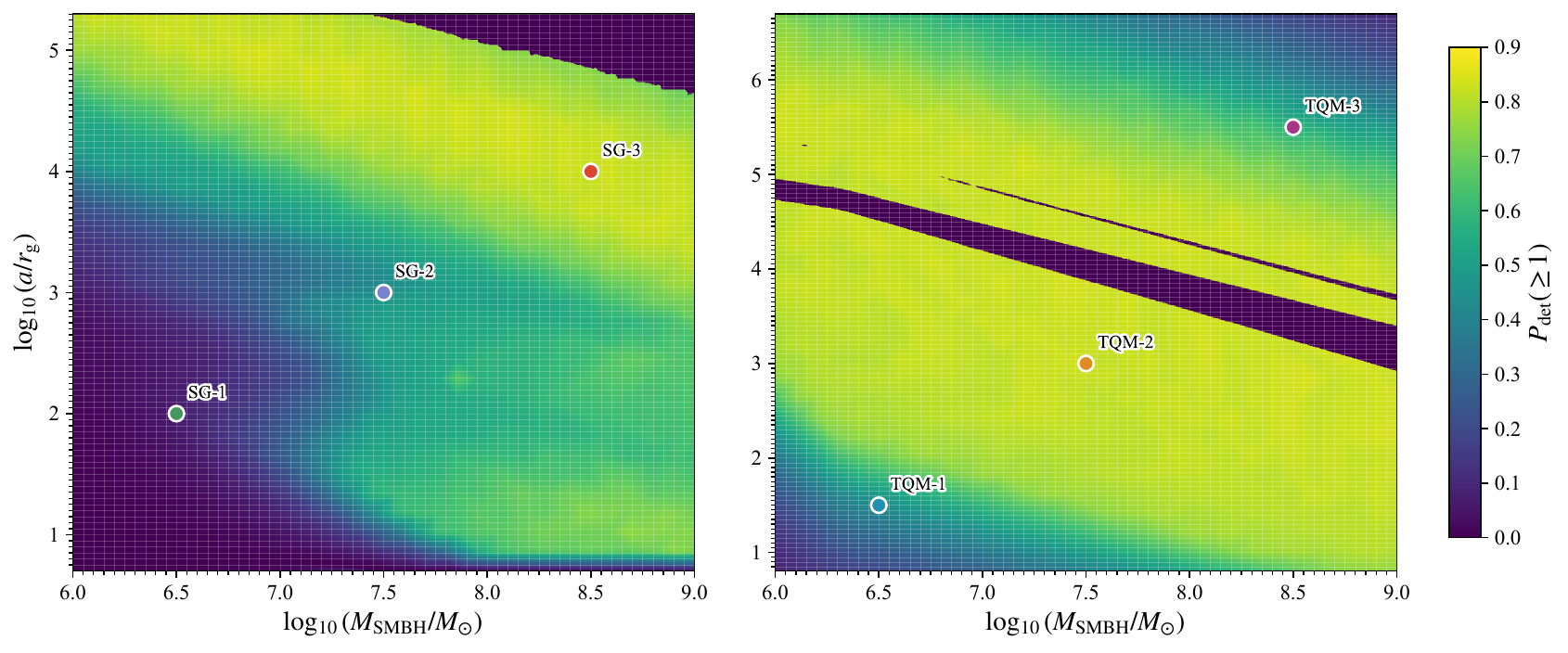}
    \caption{Parameter-space estimate of the probability of detecting at least one neutrino from a single embedded GRB jet at $D_L=100\,{\rm Mpc}$ in the SG (left) and TQM (right) disk models. Colored circles mark the six representative trajectories analyzed in section~\ref{sec:single}, styled in their corresponding case colors.}
    \label{fig:nupara_detection}
\end{figure}

\label{sub:parameter space}
Assuming Poisson statistics, the probability of detecting at least one neutrino from a single burst is~\cite{mukhopadhyay2024}
\begin{equation}
    \label{eq:probability}
P_{\rm det}=1-e^{-N_{\rm det}}.
\end{equation}
Figure~\ref{fig:nupara_detection} presents the $P_{\rm det}$ distribution across the $(M_{\rm SMBH}, a)$ parameter plane.
The maps reveal distinct structural behaviors regulated by the underlying disk models: in the SG model, the high-probability region broadens and shifts outward toward higher SMBH masses, whereas the TQM model generates an extended high-probability plateau across a wider parameter region with a weaker mass dependence. Peak detection probabilities reach $\approx 82\%$, corresponding to $N_{\rm det}\approx 1.7$. On the breakout side of the parameter plane, short engine-clock ratios ($\mathcal{R}_{\rm dur} < 1$) result in lower detection probabilities due to early dynamical truncation of embedded emission.

This behavior illustrates the impact of parameter-dependent spectral shifts: for SG-3, reduced particle cooling preserves a broad high-energy tail extending into the PeV range that still aligns with the peak of IceCube's effective area, resulting in a high detection probability ($P_{\rm det} \approx 76\%$). For TQM-3, by contrast, the weaker cooling produces a harder spectrum with an extended $\sim10^2\,\mathrm{PeV}$ tail, while its IC86 muon-track event count is lower than that of TQM-2 ($N_{\rm det} \approx 0.72$ versus $1.38$). The high-energy tail lies increasingly outside the energy range that dominates the IC86 yield.

\section{Discussion and conclusions}%
\label{sec:discussion}

In this work, we have established a time-dependent framework connecting the vertical structure of AGN accretion disk to jet-head propagation, reverse-shock particle acceleration, and high-energy neutrino radiation. By applying this model to the SG and TQM disk prescriptions, we evaluated time-resolved neutrino spectra, light curves, time-integrated fluences, and detector yield distributions across the $(M_{\rm SMBH}, a)$ parameter space.

The accuracy of a single-state approximation depends on the propagation regime. For choked jets, the head decelerates within the disk and the reverse-shock power, proportional to $1-\beta_{\rm h}$, becomes concentrated near the terminal stage of the trajectory. The stalling state therefore represents the fluence-dominant phase, rendering stalling-state approximations remarkably robust. For breakout jets, however, late-stage acceleration drives outward growth of maximum proton energy and rapid spectral reshaping, causing pre-breakout approximations to overpredict fluences even after energy truncation. Resolving full jet propagation histories is therefore physically indispensable.

Furthermore, the ambient disk density structure regulates the neutrino production, creating distinct detection landscapes across the parameter space. Both disk prescriptions favor higher detection probabilities toward larger SMBH masses and larger disk radii. The TQM model supports a broader high-probability plateau, followed by a decline in the most massive and radially extended environments. The representative low-density, deeply choked TQM-3 trajectory experiences reduced meson and muon cooling, producing a harder neutrino spectrum with a high-energy tail that reaches $\sim10^2\,\mathrm{PeV}$. This spectral hardening shifts part of the neutrino output away from the energy range that contributes most efficiently to the IC86 muon-track event yield. The high-energy tail overlaps the ultra-high-energy range targeted by the planned IceCube-Gen2 Radio Array~\cite{aartsen2021} and approaches the low-energy edge of the Giant Radio Array for Neutrino Detection (GRAND)~\cite{alvarez-muniz2020}. Both are planned to achieve large apertures at ultrahigh energies.

We include neutrinos produced at the reverse shock but not independent cocoon contributions. In the present calculation, particle acceleration is completely suppressed whenever the reverse shock satisfies the RNS condition, possible collisionless subshocks in magnetized RMSs could modify this boundary~\cite{beloborodov2017a, ai2026}. The present calculation assumes a constant top-hat engine with fixed jet parameters. Time-variable engines, multidimensional jet--cocoon dynamics, and source-to-source variations may introduce additional temporal and spectral structure.

Finally, future forecasts of the diffuse neutrino background from embedded transients must incorporate distributions of engine duration, disk structure, and compact-object merger or collapse rates. The time-dependent framework presented here provides the physical foundation required for such population-level multi-messenger synthesis.

\appendix
\section{Flavor composition}
To quantify the energy-dependent flavor composition at Earth, we define the Earth-side muon-flavor fraction as
\begin{equation}\label{eq:numu-fraction}
   f_{\nu_\mu+\bar\nu_\mu}^\oplus(E_\nu)=
\frac{F_{\nu_\mu+\bar\nu_\mu}^{\oplus}(E_\nu)}
{\sum_{\alpha=e,\mu,\tau}F_{\nu_\alpha+\bar\nu_\alpha}^{\oplus}(E_\nu)}.
\end{equation}
\begin{figure}[H]
   \centering
   \includegraphics[width=0.6\textwidth]{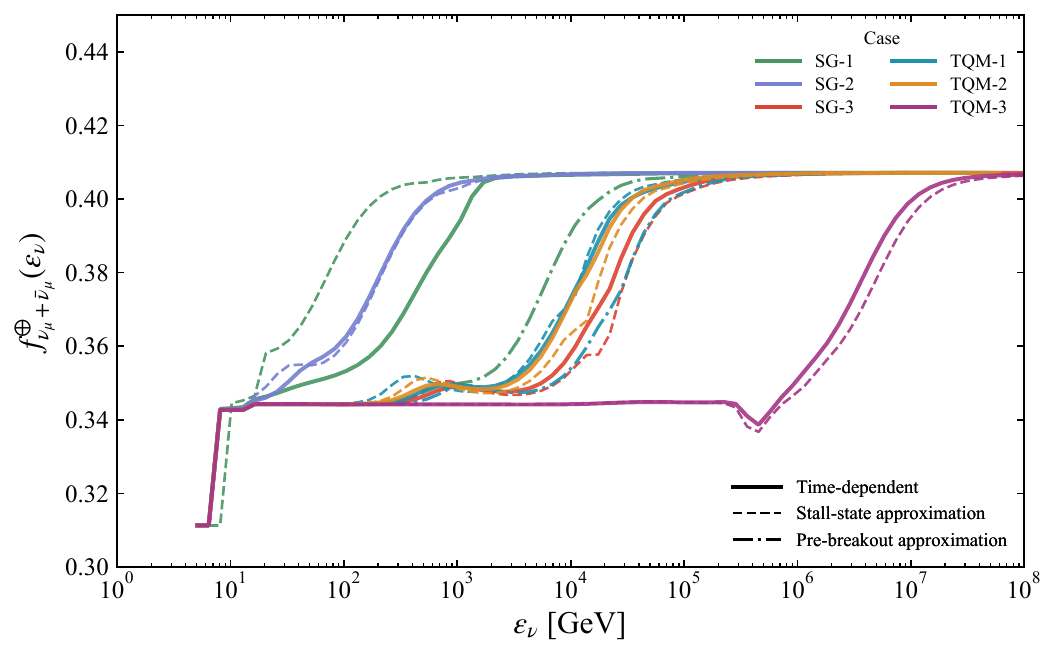}
   \caption{Muon-flavor fraction of the time-integrated neutrino fluence at Earth after oscillation-averaged flavor transitions. Solid, dashed and dash-dotted curves show the time-dependent caculation, single-state approximation and pre-breakout breakout (only for breakout cases) calculations, respectively. Colors identify the six representative cases ($\mathrm{SG}\text{--}1\dots 3$ and $\mathrm{TQM}\text{--}1\dots 3$).}
   \label{fig:numu_fraction_earth}
\end{figure}
Figure~\ref{fig:numu_fraction_earth} presents the Earth-side muon-flavor fraction $f_{\nu_\mu+\bar\nu_\mu}^\oplus(E_\nu)$ after standard PMNS oscillations~\cite{esteban2024}. Overall, oscillation-averaged flavor mixing confines the fraction to a narrow $0.34\text{--}0.41$ range across the energy spectrum. Comparing the time-dependent results with the both single-state approximations reveals a prominent discrepancy for SG-1, whereas the choked trajectories and TQM-1 breakout case exhibit more similar flavor profiles. 

\acknowledgments
We acknowledge Hao-Hui Zhang for useful discussions and Jin-Ping Zhu for providing the code used to model the structures of the SG and TQM disks. This work is supported by the National Key R\&D Program of China (2021YFA0718500) and the National Natural Science Foundation of China (grant No. 12393811).



\bibliographystyle{JHEP}
\bibliography{ref}

\end{document}

%% file: pic/disk_parameters_sixproto.tex
{
    \small
    \setlength{\tabcolsep}{4.5pt}
    \begin{tabular}{lcccccc}
    \toprule
    Case & $(\log_{10}\frac{M_{\rm BH}}{M_\odot},\,\log_{10}\frac{a}{r_{\rm g}})$ & $\rho_0\,[\mathrm{g\,cm^{-3}}]$ & $\kappa\,[\mathrm{cm^{2}\,g^{-1}}]$ & $H\,[\mathrm{cm}]$ & $\mathcal{R}_{\rm dur}$ \\
    \midrule
    SG-1 & $(6.5,\,2)$ & $8.118\times10^{-8}$ & $0.3472$ & $1.205\times10^{12}$ & $0.273$ \\
    SG-2 & $(7.5,\,3)$ & $1.230\times10^{-8}$ & $0.6031$ & $3.521\times10^{13}$ & $53.9$ \\
    SG-3 & $(8.5,\,4)$ & $9.832\times10^{-13}$ & $0.3729$ & $1.220\times10^{16}$ & $4.41\times10^{4}$ \\
    TQM-1 & $(6.5,\,1.5)$ & $7.425\times10^{-12}$ & $0.2021$ & $2.123\times10^{13}$ & $0.606$ \\
    TQM-2 & $(7.5,\,3)$ & $3.178\times10^{-12}$ & $0.3183$ & $1.111\times10^{14}$ & $12.8$ \\
    TQM-3 & $(8.5,\,5.5)$ & $5.954\times10^{-17}$ & $2.8680$ & $6.441\times10^{17}$ & $9.85\times10^{5}$ \\
    \bottomrule
    \end{tabular}
    }

%% file: pic/six_case_results.tex
\begin{table}[tbp]
    \centering
{    \small
    \setlength{\tabcolsep}{4.5pt}
    \begin{tabular}{lrrrrrr}
        \toprule
        Case & \multicolumn{2}{c}{TD} & \multicolumn{2}{c}{Stalling-State} & \multicolumn{2}{c}{Pre-Breakout} \\
        & $N_{\rm det}$ & $P$ & $N_{\rm det}$ & $P$ & $N_{\rm det}$ & $P$ \\
        \cmidrule(lr){2-3}\cmidrule(lr){4-5}\cmidrule(lr){6-7}
        \midrule
        SG-1  & 0.104 & 0.098 & 0.283 & 0.247 & 0.267 & 0.234 \\
        SG-2  & 0.443 & 0.358 & 0.438 & 0.355 & -- & -- \\
        SG-3  & 1.418 & 0.758 & 1.507 & 0.778 & -- & -- \\
        TQM-1 & 0.782 & 0.542 & 1.401 & 0.754 & 0.849 & 0.572 \\
        TQM-2 & 1.380 & 0.748 & 1.479 & 0.772 & -- & -- \\
        TQM-3 & 0.722 & 0.514 & 0.837 & 0.567 & -- & -- \\
        \bottomrule
    \end{tabular}}
    \caption{Detection summary for the six representative trajectories at $D_L=100\,\mathrm{Mpc}$. The time-dependent (TD) and stalling-state columns give the expected IC86 muon-track count $N_{\rm det}$ and detection probability $P=1-\exp(-N_{\rm det})$. For the two breakout cases, the pre-breakout columns evaluate the single-state approximation just before breakout using energy normalization $E_{\rm iso}t_{\rm req}/t_{\rm dur}$; dashes denote choked cases for which this baseline is undefined.}
    \label{tab:six_case_results}
\end{table}